\documentclass[12pt,preprint]{aastex}

\def\gtorder{\mathrel{\raise.3ex\hbox{$>$}\mkern-14mu
             \lower0.6ex\hbox{$\sim$}}}
\def\ltorder{\mathrel{\raise.3ex\hbox{$<$}\mkern-14mu
             \lower0.6ex\hbox{$\sim$}}}

\def\Msun{\>{\rm M_{\odot}}}
\def\Rsun{\>{\rm R_{\odot}}}

\shorttitle{M dwarf Wind Measurement}
\shortauthors{Debes}

\begin{document}
\title{Measuring M dwarf Winds with DAZ White Dwarfs\footnote{Based on observations made with the NASA/ESA Hubble Space Telescope, obtained from the Data Archive at the Space Telescope Science Institute, which is operated by the Association of Universities for Research in Astronomy, Inc., under NASA contract NAS 5-26555. These observations are associated with program 10255}}
\author{John H. Debes\altaffilmark{2}}

\altaffiltext{2}{Department of Terrestrial Magnetism, Carnegie Institution of Washington, Washington, DC 20015, debes@dtm.ciw.edu}

\begin{abstract}
Hydrogen atmosphere white dwarfs with metal lines, so-called DAZs,
 show evidence for ongoing accretion of material onto their 
surfaces.  Some DAZs are known to 
have unresolved M dwarf companions, which could
account
for the observed accretion through a stellar wind.  I combine observed Ca 
abundances of the DAZs with information on the orbital separation of their 
M dwarf companions to infer the mass loss rate of the M dwarfs.  I find that
for three of the six known DAZs with 
M dwarf companions, a stellar wind can 
plausibly explain the observed accretion on the white dwarfs assuming Bondi-Hoyle accretion of solar abundance stellar winds on the order of 10$^{-14}$-10$^{-16}\Msun$\ yr$^{-1}$.  The rest of the sample have companions with
orbits $\gtorder$ 1~AU, and require companion mass loss rates of $> 10^{-11}\Msun$\ yr$^{-1}$.  I conclude that there must be an alternative explanation 
for accretion of material onto DAZs with widely separated companions.  
The inferred winds for two of the close
binaries are orders of magnitude smaller than typically 
assumed for the angular momentum
loss of red dwarf-white dwarf pairs due to magnetic braking from a stellar wind
and may seriously affect predictions for the formation rate of CVs with low
mass companions.
\end{abstract}   


\section{Introduction}
Hydrogen atmosphere white dwarfs with spectral absorption lines due to metals 
are a potential puzzle for white dwarf evolution.  Metals present in the
upper atmossheres of white dwarfs
with T$_{eff}$ $<$ 20.000 K generally diffuse out of the upper atmosphere on
timescales ranging from days to 10$^6$ yr \citep{paquette86}.  It was therefore
expected that white dwarfs with cooling ages $\gg$10$^{8}$~yr should show no detectable metal line absorption.  However a number of both hydrogen (class DA)
and helium (class DB) white dwarfs show the evidence of heavy metals in their
 atmospheres (classes DAZ and DZ).
The initial explanation for metal contamination was the accretion of solar 
abundance ISM material\citep{dupuis93b,dupuis92,dupuis93a}.  But the lack of 
correlation between white dwarfs and dense ISM clouds, a large number of
DAZs residing at $<$70 pc within the Local Bubble, the high 
inferred accretion rates for DAZs, and the lack of hydrogen in DZ atmospheres
 makes ISM accretion problematic\citep{aannestad93,zuckerman98,zuckerman03}.
Recent surveys of white dwarfs have turned up several new DAZs, and these have
been used to put forth a separate ISM accretion argument for within the Local Bubble \citep{koester05,koester06}.  In this case, nearby white dwarfs accrete
continuously from clouds of warm, partially ionized ISM

Other explanations exist, however.  The direct
accretion of heavy metals through cometary impacts with white dwarfs was
invoked to explain the first DAZ, G74-7 \citep{alcock86}.  This explanation 
can deliver a moderate amount of material to the white dwarf surface, but fails
to explain high abundance DAZs with short settling times and dusty disks
such as G29-38 and GD 362 unless a planetary
system perturbs a steady supply of material close to the white dwarf 
\citep{zuckerman87,koester97,holberg97,debes02,gianninas04}.
  
In a survey of DAs for metal lines, six of ten known unresolved
DA+M dwarf systems were DAZs compared to 1/9 for double degenerates 
and 0/7 for resolved
binaries\citep{zuckerman03}.  This raises the possiblity that DAZs may be 
caused by hitherto unseen stellar or substellar companions that deposit material through a stellar wind onto the surfaces of the white dwarfs.  The evidence
for apparently single DAZs with unknown companions is thin.  Only two known substellar companions
to white dwarfs exist, in widely separated binaries \citep{zuckerman92,farihi04}.  This is despite wide searches for unresolved stellar
or substellar objects in close orbits \citep{probst82,zuckerman92,dobbie05,farihi05}.  Furthermore, concerted searches of DAZs 
themselves show that stellar companions are ruled out at all separations 
and substellar companions ruled out for all but the closest orbital separations
\citep{debes05a,debes05b,debes05c}.  

M Dwarf stellar wind mass loss 
rates are poorly understood and hard to study--upper
limits can only be placed on the nearby M5.5 
Proxima Centauri based on X-ray and Ly $\alpha$
studies with upper limits of 6$\times 10^{-14}M_{\odot}$/yr (3 $\dot{M}_\odot$)
 and 4$\times 10^{-15} M_\odot$/yr (0.2 $\dot{M}_{\odot}$) 
respectively \citep{wargelin02,wood01}.  Radio observations with marginal
detections and spectroscopic 
observations of coronal mass ejections from low mass stars have been used
to propose very high mass loss rates from dwarf stars, which has a theoretical
motivation \citep{mullan92,foing90,badalyan92}.  These results are 
controversial, however, and depend on the model assumptions made \citep{lim96}.
The recent results for Proxima Centauri suggest that perhaps M dwarf stars have
winds comparable to or smaller than solar type stars, rather than the larger 
rates.  The strength of young M dwarf winds can affect the lifetime
of gaseous material in disks that are important for planet formation
and limit the presence of
dusty disks \citep{laughlin05,plavchan05}.  In orbit around white dwarfs,
M dwarf winds are important for the origin of cataclysmic variables
\citep{g03}.

Under the assumption that the cause of the metal lines in DAZs is due to
M dwarf companions, the stellar wind rates for known DAZ+M dwarf systems
can be calculated.  These calculations require the known orbits of the M 
dwarf companions.  Three of the known DAZ+M systems have well known
orbital periods and masses since they are detached transiting systems.
The other three have recent HST Advanced Camera for Surveys (ACS) images
that allow orbital information to be estimated.

In this paper I infer the stellar wind mass loss rates of the M dwarf companions to six DAZ+M dwarf binaries.   In Section \ref{s1} I estimate the orbital
semi-major axis for each binary.  In Section \ref{s2} I use the semi-major axes and white dwarf
diffusion coefficients to infer the mass accretion rate onto the white dwarfs
and thus the mass loss rate of the companion M dwarfs.  Finally, in Section \ref{s3} I discuss my results.

\section{Determining Orbital Separations}
\label{s1}
\subsection{Close Binaries}

Three of the six DAZs with companions have previously measured orbital periods,
primary, and companion masses.  WD 0419-487 is a 0.47$\Msun$\ white dwarf with a 0.095
companion in a 7.3 hr orbit.  The spectroscopic survey of DAs
that discovered WD 0419-487 noted that there was evidence of a hot 
spot/accretion disk in the observed spectrum \citep{zuckerman03}.  I would
expect that this object may show a higher accretion rate due to some amount of
mass transfer in addition to that caused by a wind.  WD 1026+002 is a 0.68 $\Msun$ white dwarf with a 0.23 $\Msun$ companion in a 14.3 hr orbit.  Finally, WD 1213+528 is a 0.63 $\Msun$ white dwarf with a 0.36 $\Msun$ companion in a  16 hr orbit.  
Table \ref{tab:closeorb} lists the orbital parameters important for my calculations.

\subsection{Widely Separated Binaries}
The remaining three systems have had their radial velocities measured to an
accuracy within a few km/s and show either slow RV trends or none at all \citep{schultz96}.  These WDs are prime candidates for high spatial resolution imaging in order to determine what their approximate orbital separations are.  Recently, a large snapshot survey (Program 10255, PI D. Hoard) with the Hubble Space Telescope's Advanced Camera
for Surveys (ACS) was performed on a large sample of previously unresolved
white dwarf+red dwarf pairs, including the three
DAZ+M systems that have no orbital information \citep[for first results, see][]{farihi06}.  
Of these three, WD 1049+103 and WD 
1210+464 show resolved companions with the ACS data.  The third DAZ, 
WD 0354+463
shows no obvious resolved companion \citep{farihi06}. 

\begin{figure}
\plottwo{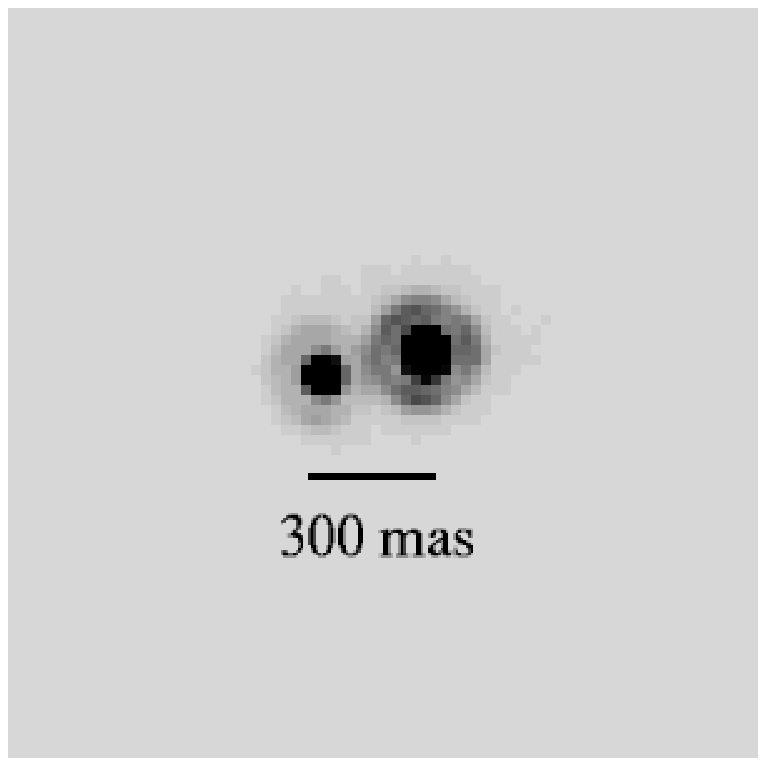}{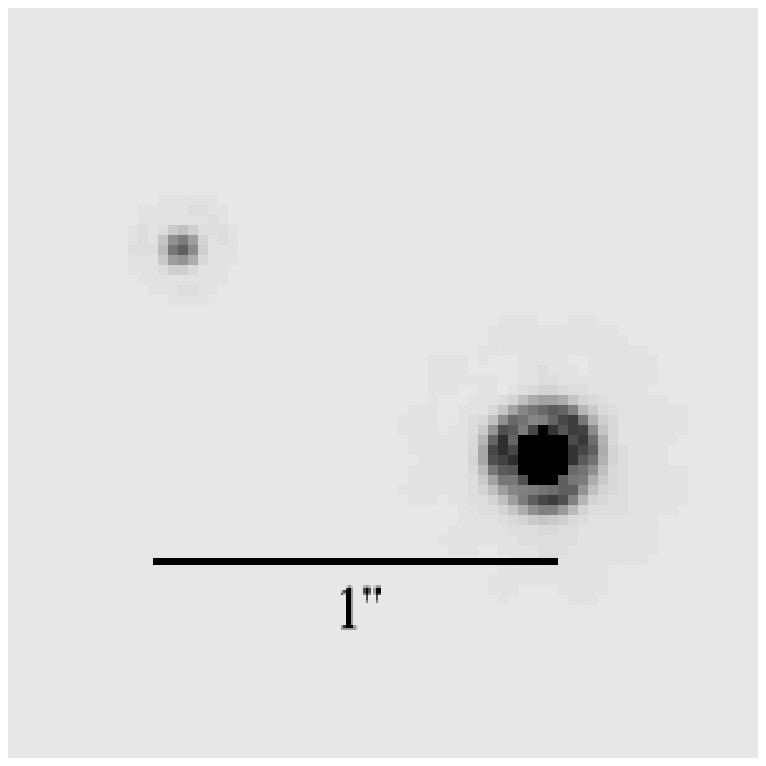}
\caption{\label{fig:acs}(left)ACS HRC image of WD 1049+103 with its resolved
M dwarf companion. (right) ACS HRC image of WD 1210+464 with its resolved 
M dwarf companion.}
\end{figure}

The STScI pipeline calibrated
ACS data for these three WDs was retrieved and analyzed by measuring the
centroid positions of the white dwarf and resolved companion.  Figure
\ref{fig:acs} shows the two imaged targets with resolved companions.
  The data was
taken with the High Resolution Camera (HRC) on ACS and so the plate scale for
the final geometrically corrected images is $\sim$0\farcs025/pixel.  The
projected radius was calculated and this was used to estimate an orbital 
separation assuming a circular orbit.  

\begin{figure}
\plottwo{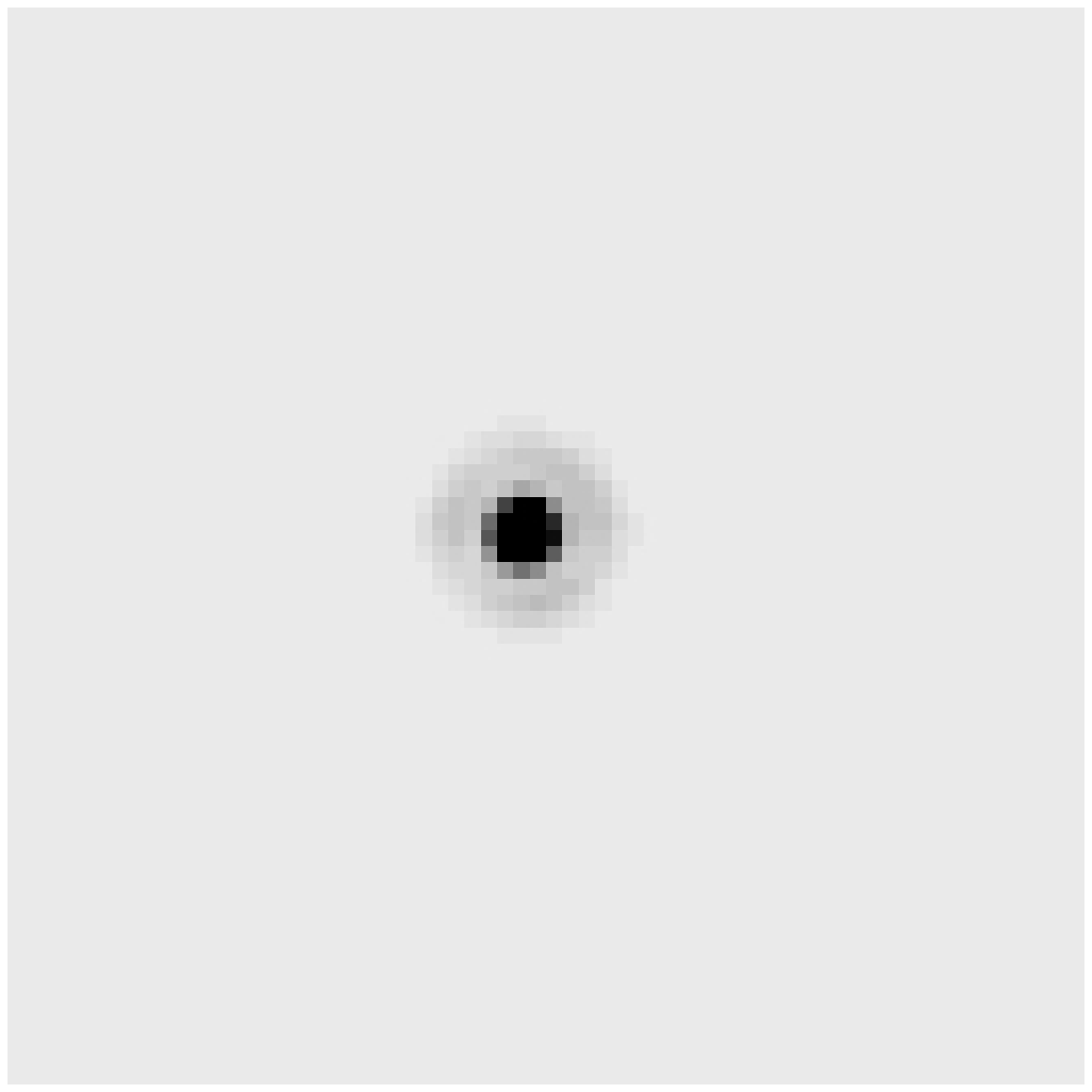}{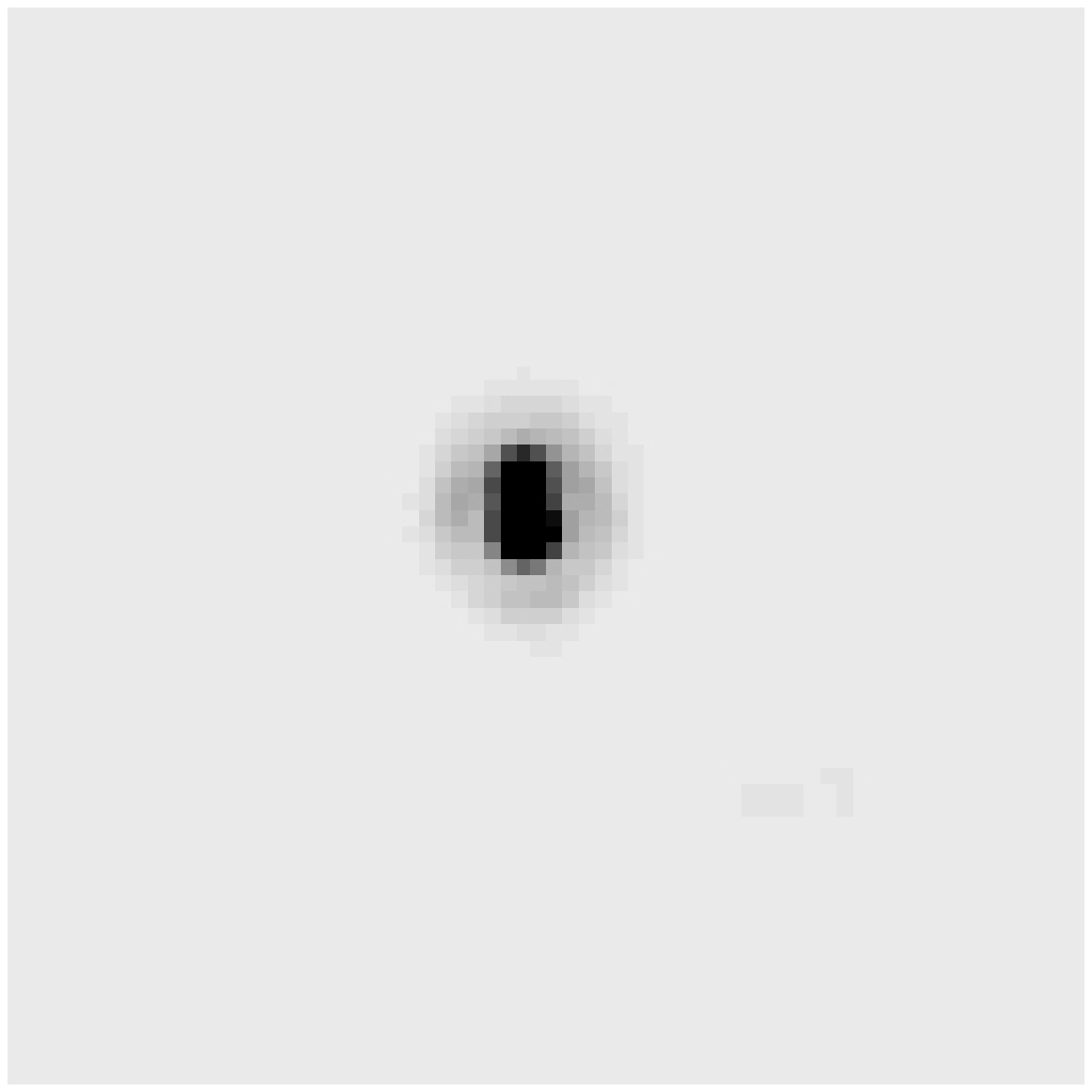}
\caption{\label{fig:sens}(left) ACS HRC image of WD 0354+463 without any
artificial implants.  (right) Same image, but with an artificial companion
with m$_{F814W}$=15.4 at a separation of 70 mas.}
\end{figure}

Since no resolved companion was detected for WD 0354+463, an estimate for
the upper limit to the M dwarf orbital separation must be done. One can
estimate the sensitivity of an image to a resolved companion by implanting
artificial companions into the data and measure when they are recovered.
I constructed a reference HRC PSF from the image of WD 1210+464.
At separations $>$70 mas or 2~AU, an 
object with m$_{F814W}$=15.4 could have been 
detected, corresponding to the approximate F814W magnitude expected for
both WD 0354+463 and the M dwarf assuming a mass of 0.1$\Msun$\citep{baraffe98}.  Figure \ref{fig:sens} shows the limiting separation at which
WD 0354+463 would have been observably separated from its companion. 

 As a lower limit to the orbital separation, I use the 
timescale of radial velocity observations by \citet{schultz96}, which
is $\sim$0.7~yr.  Assuming a
primary white dwarf mass of 0.6$\Msun$, I calculated a lower bound on the
orbital separation based on no observed variations.  If I assume that  
 at least half a period would have been detected and neglect the mass of the companion, I obtain a lower bound on the orbital semi-major axis of $\sim$1~AU.
The projected separation would be $\sim$35~mas, well below the spatial 
resolution of ACS.  I adopt a value of 1.5$\pm$0.5~AU for the orbital
separation of the M dwarf companion.

Finally, one cannot estimate an orbital separation without knowledge of the
white dwarf's distance from the Sun.  None of these three
objects have parallax measurements so I utilize the
determinations of the distance based on model fits to broadband photometry.
WD 0354+463's distance can be calculated by an estimate of its absolute magnitude, which gives a distance of $\sim$30~pc \citep{sion}.  For WD 1049+103, and
WD 1210+464,
both objects are part of the PG survey, which has been modeled by \citet{liebert04}.  The model distances and the estimated semi-major axes are listed in
Table \ref{tab:wideorb}.

\section{Inferring Mass Accretion and Stellar Wind Rates}
\label{s2}
Inferring the mass accretion onto the white dwarf proceeds from a knowledge of
the number abundance of some tracer element relative to hydrogen.  In the case
of DAZs the strongest observed lines in the visible 
tend to be the Ca H and K lines.  If one assumes a solar composition
of material accreted, the accretion rate is then
\begin{equation}
\label{eq:mdot}
\dot{M}=\frac{q M_{wd}[Ca/H]}{\Theta [Ca/H]_{\odot}}
\end{equation}
where $q$ is the mass fraction at the base of the convection zone and $\Theta$ is the e-folding timescale for Ca 
to settle out of the atmosphere \citep{paquette86}.  [Ca/H]$_\odot$
is the solar value of [Ca/H] or 2.1875$\times$10$^{-6}$.  I use the observed
number abundances for the DAZs
calculated in \citet{zuckerman03}.  For WD 1210+464, there is
no abundance calculated, but based on the upper limits based on non-detections
of other white dwarfs in that survey, I estimate that [Ca/H] $\sim$-7.
 I discuss this further in Section \ref{s3}.  It should be noted that based
on its T$_{eff}$, WD 1210+464 may be hot enough to sustain a radiative wind which would create metal lines, 
thus making it irrelevent that it has a companion.  For completeness, I still
calculate the inferred mass loss rate of its companion.

For this paper, I use the values of $q$ determined by \citet{althaus98} and the values of $\Theta$ by \citet{paquette86}, even though these two results use
different methods of calculating turbulence.  This method has been shown to 
produce the accretion rate for the DAZ G29-38 reasonably well in light of independent estimations of the accretion rate \citep{graham90,debes02}.  Additionally for WD 1049+103 and WD 1210+464, their T$_{eff}$ are higher than those 
calculated by either method, so I use the values from the highest effective
temperatures calculated.

Inferring the mass loss rate of the companion requires a mechanism 
for accretion.  For simplicity and without knowledge of the exact mechanism for
creating a stellar wind, I assume the M dwarf companion expels a spherically
symmetric flow at the escape speed from the surface of the M dwarf.  I assume
that the white dwarf accretes material through a Bondi-Hoyle 
accretion flow determined by
\begin{equation}
\label{eq:bondi}
\dot{M}=\frac{4\pi G^2 M_{WD}^{2}\rho(R)}{v_{rel}^3}
\end{equation}
where $G$ is the gravitational constant, $\rho$ is the density of material surrounding the WD, and $v$ is the relative velocity to the WD at which the gas is
passing, which I take to be $\sqrt{v_{wind}^2+v_{orb}^2}$ \citep{boni}.
The escape speed from a low mass star is $\sim$600 km/s, assuming that $M/R\sim1$.  It is possible that
the widely separated binaries have a higher mass than the close binaries.  The mass-radius relation is roughly true for most low mass stars as can be seen in Table \ref{tab:closeorb} \citep[but see][]{lopez05}.  
However, WD~0419-487's radius is roughly twice as large as would be expected.  This could be related to the possiblitity of its red dwarf undergoing mass transfer with the white dwarf, or the fact that it 
is most likely tidally locked and is a fast rotator.  In any case, the assumption of 600 km/s for an escape speed for the widely separated binaries I'm
studying is probably safe, since they should be similar to field red dwarfs.

Using the escape speed at the 
radius of the companion may not be strictly true.  For example, one could
model isothermal or polytropic winds from the M dwarfs assuming some heating
mechanism.  This is highly model dependent and requires information about the
M dwarf companions that is not easily determined by the
observations at hand.  Presumably the main source
of heating for the companions is comparable to that for the solar wind, where
coronal heating provides the bulk of energy for the acceleration of the wind.
Uncertainties in $v$ are discussed further in Section \ref{s3}.

Additionally, Bondi-Hoyle accretion may not accurately describe these 
systems.  Many
astrophysical objects show departures from the simple Bondi-Hoyle picture, such
as isolated neutron stars accreting from the ISM \citep{perna03} and
super massive black holes \citep{dimatteo01}.  In these cases, there are departures from
the simple plane parallel geometry of the Bondi-Hoyle case due to turbulence,
and magnetic fields \citep{krumholz05}.  These departures serve to suppress
the Bondi Hoyle accretion rate, making accretion more inefficient.  Wind accretion in binary systems as well can depart from the simple Bondi-Hoyle case, 
particularly when $v_{orb}\sim v_{wind}$.  In these cases, the accretion rates
can be an order of magnitude lower \citep{theuns93,theuns96}.  For all of these
situations, the accretion supression results in higher inferred
wind mass loss rates.  The values I calculate correspond to strict lower
limits if simple Bondi-Hoyle accretion is not present.

The final step is to link the mass loss rate of the red dwarf to the observed
accretion rate onto the WD.  The density depends directly on this mass loss. 
I assume that the mass loss rate is constant and calculate the density due to 
the continuity equation \citep{lamers}:
\begin{equation}
\rho(R)=\frac{\dot{M}_{RD}}{4\pi v R^2}
\end{equation}
where $\dot{M}_{RD}$ is the mass loss rate.  The equation can be combined 
with Equations \ref{eq:mdot} and \ref{eq:bondi} and solved for $\dot{M}_{RD}$ in terms of known 
or inferred quantities:
\begin{equation}
\label{eqn:mrd}
\dot{M}_{RD}=\frac{q[Ca/H]}{\Theta[Ca/H]_\odot}\frac{R^2 v^4}{G^2 M_{WD}}
\end{equation}

I tabulate my results for the known close binaries and the resolved binaries in Table \ref{resulttab}.  For comparison, the solar wind mass loss rate is 
$\sim2\times10^{-14}$ M$_\odot$/yr.

The uncertainties in Equation \ref{eqn:mrd} can be roughly quantified.
The close binaries have more accurate measures of $\dot{M}$, since
M$_{WD}$,$R$, and [Ca/H] are all known to within a few percent.  In their cases
the formal statistical error based on these measurements comes to $<$10\%.
For the widely separated binaries, the errors
in M$_{WD}$ and [Ca/H] are on the order of $\sim$10\% as well.  
Assuming that
the model parameters and $v$ are correct, $\dot{M}_{RD}$ is well
constrained. 

Since the results are model dependent, systematic errors most certainly dominate over the statistical errors estimated above.  
Uncertainty is present in the calculation of $\Theta$, which
depends on $q$, the mass fraction of the convective envolope,
the diffusion coefficients used, and of the 
convective envelope, which is highly sensitive to certain ranges of temperature
and depends on the formalism of turbulence chosen \citep[see][for example]{althaus98}.  At T$_{eff}$ $>$12000 K and $<$ 8000 K, the various flavors of 
turbulence tend to converge within factors of a few, while between these two
ranges values for $q$ vary by up to four orders of magnitude.  This uncertainty is helped somewhat by
the fact that $\Theta \propto q^{0.7}$\citep{paquette86}.
  Equation \ref{eqn:mrd} shows that
$\dot{M}_{RD}\propto q^{0.3}$.  Thus an uncertainty of up to four orders of magnitude in $q$ corresponds to a uncertainty in $\dot{M}$ of a factor of 15.  To self-consistently calculate $q$ and $\Theta$, the same turbulence formalism should be employed to determine both quantities.  

Unfortunately, a self-consistent calculation of the diffusion times
using different turbulences has yet to be synthesized
properly for the range of temperatures, masses, and atmospheric compositions
relevant to cool DAZs, and understanding of the underlying mechanism
for forming DAZs would be greatly advanced by an update of this important work
\citep[see also][]{zuckerman03}.  Most of the DAZs in my sample are in the 
range where the various models converge well to within factors of two, so
the uncertainty in $q/\Theta\sim$25\%.

The distance $R$, determined from the estimated orbital separations can be 
uncertain for the 
widely separated binaries.  Since these are projected distances, they do not necessarily correspond to the true orbital separation of the pair.
However, any such uncertainty would serve to raise the inferred stellar wind
mass loss rate since the projected orbital separation would always be smaller
than the true orbital separation.  Conversely, the companions could be in high
eccentricity orbits which brought the star much closer than observed.  Such a
situation would mean that the mass loss rate would be overestimated.  In that
case, $R$ would have to change by a factor of three (corresponding to an
orbital $e\sim$0.5) to roughly produce an order of magnitude change in
$\dot{M}_{RD}$.  It is not guaranteed that the resolved companions are 
physically bound
since there is no common proper motion information--the observations were
taken at one epoch.  The extremely close separations ($<$1\arcsec)
and the fact that the blended near-IR photometry are consistent with companions
at the same distance makes chance alignments unlikely.  

The solar wind velocity
is variable and ranges from 60\% to greater than the escape speed of the Sun, implying $v$ could have a range of values that probably does not exceed
a factor of two.  The true velocity of the wind from an M dwarf depends 
sensitively on the model for that wind, which is beyond the scope of this 
paper.  In the case of the solar wind, the measured velocity at 1~AU is 
less than that predicted for an isothermal or polytropic wind \citep{lamers}.
  Since the inferred mass loss rate is $\propto v^4$, a change
in estimated speed by a factor of two
 would change the calculated mass loss rates
by an order of magnitude.

To sum up, systematic errors in $\dot{M}_{RD}\propto q^{0.3}R^2v^4$, with $v$
creating the largest systematic uncertainty for the close binaries and $R$
and $v$ creating the largest uncertainties for the resolved systems.  The 
estimates presented in Table \ref{resulttab} are accurate to an order of
magnitude for the close binaries while the resolved binaries are accurate to 
within two orders of magnitude.  This is comparable to the range of upper 
limits given for the mass loss rate of Proxima Centauri, which approximately
spans an order of magnitude.

\section{Discussion}
\label{s3}
Our results break down along the lines of those DAZs with known close companions (a$<$0.02 AU) and those with greater separations.  At smaller separations,
the calculated mass loss rates agree reasonably well with the upper limit to
the wind around Proxima Centauri \citep{wargelin02,wood02}.  The mass loss rates I calculate
are about two orders of 
magnitude smaller than Proxima Centauri's smaller upper limit, with the
exception of WD 0419-487 which is comparable, though larger than the other
two M dwarfs.  This larger rate could be explained by moderate
Roche lobe overflow, evaporation of the companion by the DAZ, or efficient capture of the companion's wind by a magnetic field.  If WD 0419-487 was efficiently capturing all of its companion's wind, $\dot{M}_{RD}$ would fall nicely in with those observed for the other two close binaries.

On the
other hand, the mass loss rates determined for the widely separated companions
are three to four 
orders of magnitude larger than the Solar wind.  This is despite
a slightly higher uncertainty of the accretion rate onto the white dwarfs.  
Most of these uncertainties would conspire to create a
higher accretion rate.  For WD 1210+464, a lower accretion rate is possible
if the detected equivalent width corresponds to a lower abundance than
assumed.  However, even at the smallest lower limit of the \citet{zuckerman03}
survey ([Ca/H]$\sim$12.8), the inferred mass loss rate of the companion
would be equivalent to the Solar Wind and two orders of magnitude higher
than the close binaries.  It is 
possible that these systems are hierarchical triples with companions undetected
by radial velocity observations but in orbits similar to the 
close binaries.  WD 1210+464 and WD1049+103 have F814W
photometry consistent with single DAZs, neglecting their resolved companions.
This strongly argues that any further unresolved companions would have to be 
quite dim and of low mass.
Conversely, M dwarfs
could have the super solar rates predicted by the earlier 
results, but in light of the estimated winds of Proxima Centauri and the
 three close M dwarfs this seems unlikely.  Furthermore, given the inferred
total ages of the host white dwarfs, the companions would have either completely evaporated or lost a large fraction of their total mass.  

The low mass loss rates for the three closest binaries has two possible interpretations.  Either the mechanism for accretion is suppressed relative to Bondi-Hoyle accretion by several orders of magnitude if one expects M dwarf winds to
be similar to the Sun, or M dwarf winds are quenched even in situations where 
they are rotating quickly and should have significant activity due to 
strong magnetic fields.  Some evidence for the quenching of
 winds for low mass stars comes from \citet{mohanty03}, who find that very
late spectral type stars have lower indicators of activity due to a corona or
chromosphere.
This has been noted in studies of the angular momentum evolution of 
CVs, where fully convective companions were believed to have lost less angular
momentum due to an inefficient dynamo process \citep{durney93}.

If the winds from these low mass companions are orders of magnitdue smaller
than the solar wind mass loss rate, this has significant implications for the
angular momentum evolution of CV progenitors and CVs themselves.  Typically,
it has been assumed that the angular momentum evolution of a CV or CV-progenitor occurred due to magnetic braking through a stellar wind, with angular 
momentum loss:

\begin{eqnarray}
\label{eq:ang}
\dot{J} & = & -K_w \left(\frac{R_{sec}}{\Rsun}\right)^{2-N}\left(\frac{M_{sec}}{\Msun}\right)^{-N/3} \left(\frac{\dot{M}}{10^{-14} \Msun s^{-1}}\right)^{1-2N/3}\omega^3\  \mbox{for}\ \omega \le \omega_{crit} \\
        & = & -K_w \left(\frac{R_{sec}}{\Rsun}\right)^{2-N}\left(\frac{M_{sec}}{\Msun}\right)^{-N/3} \left(\frac{\dot{M}}{10^{-14} \Msun s^{-1}}\right)^{1-2N/3}\omega \omega_{crit}^2\ \mbox{for}\ \omega > \omega_{crit} \\
\end{eqnarray}
where $\omega_{crit}$ represents a cutoff rotation speed where the magnetic field saturates.
This equation is a modified form of that used by \citet{kawaler88}, following the work
of \citet{mestel87} and \citet{weber67}.  This prescription was used by
others to estimate the timescale for the evolution of single low mass 
stars, the onset of mass transfer in CV progenitors, and the timescale for the evolution of CVs \citep{sills00, g03, andronov03}.  In all of these cases, it 
was implicitly assumed that the mass loss rate was approximately
the solar wind mass loss rate.  $K_w$ and $N$ were essentially free parameters
 but were chosen to be 2.7$\times$10$^{47}$ g cm s and 1.5 respectively in these works.  $K_w$ essentially collects all the uncertainties in the properties of the wind and 
magnetic field of a particular star and $N$ corresponds to the particular 
magnetic field geometry.  It is immediately apparent that for the given $N$,
$\dot{J}\propto\dot{M}$ and the consequent angular momentum loss rate for the
observed DAZs would be 1-2 orders of magnitude smaller than would be calculated
by Equation \ref{eq:ang}.  The consequent timescale for these binaries to become CVs would be dominated by gravitational radiation, and would not be affected
significantly by magnetic braking.

\begin{figure}
\plotone{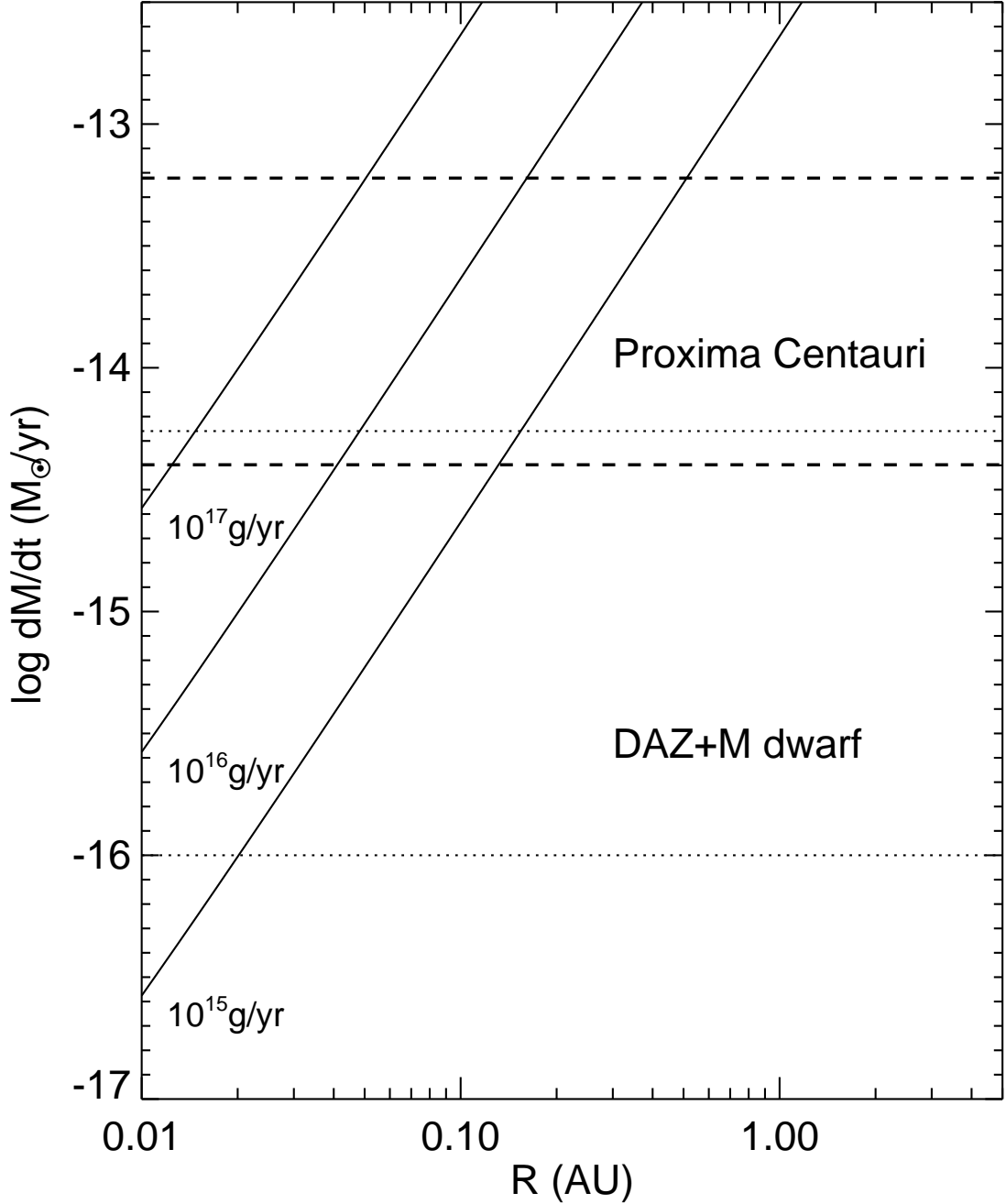}
\caption{\label{fig:cutoff}Calculation of stellar mass loss rates for 
given observed white dwarf accretion rates as a function of distance.  
Each curve corresponds to a different level of accretion one would observe
on a white dwarf. The horizontal dashed lines encompass the range of mass
loss rate values observed as upper limits for Proxima Centauri and the dash-dotted lines encompass the rates calculated for the DAZ+M dwarf sample}
\end{figure}

I also investigate a plausible orbital separation cutoff where a typical measured
WD accretion rate can be maintained by an M dwarf wind.  Using a similar calculation as I used for my widely separated binaries, 
I place a cutoff for reasonable
mass loss rates due to winds at the stellar wind upper limits for
Proxima Centauri (See Table \ref{resulttab}).  
Figure \ref{fig:cutoff} shows the result for varying levels of 
mass accretion that should be typical for DAZs.  The lowest accretion
rates allow plausible companions at wider separations, but the maximum
allowed is $\sim$1~AU.

It may be instructive to separate DAZs into three classes for ease of identifying
the underlying mechanism for accretion.  The first class would be DAZs with
companions in orbits $\ltorder$1~AU.  These are most plausibly explained by
a wind scenario and can be used to measure M dwarf winds to rates
orders of magnitude smaller
 than other methods \citep{wood01,wood02}.  The second class would encompass
all singular WDs and WDs in binaries $\gtorder$1~AU, with distances from the
Sun $<$70~pc.  These DAZs cannot easily be explained by either ISM accretion
or the companion wind explanation.  The third class of DAZs would be
at distances $\gtorder$70~pc and be either singular or have companions at separations $\gtorder$1~AU.  These objects would either have to show ISM accretion through correlation with a known dense cloud or through spectroscopic evidence
of a surrounding medium.  Otherwise they would have the same origin as those in
the second class.

ISM accretion remains viable for the two binaries that are at larger
distances, though one would expect these two objects to correspond with known
clouds or show some other evidence of a fairly dense ISM component.
WD~0354+463, however, is well within the
Local Bubble. Both wind accretion and ISM accretion would have difficulty explaining this object.  Planetesimal accretion could still be possible.  
This can either
occur through comet impacts and disruptions such as suggested by
\citet{alcock86} and \citet{debes02}, or through the tidal disruption of 
asteroids as suggested by \citet{jura03}. Relic planetesimals from a circumbinary ring around the two stars could
provide the necessary materials.  The M dwarf could be slowly perturbing the
planetesimals. Since extrasolar planets are known to exist in
binary systems, the interaction between a planet and M dwarf could be 
perturbing planetesimals into WD crossing orbits.  If this is the case, the resevoir of planetesimals might be detectable with {\em Spitzer} or at longer wavelengths.

Finally, there are roughly 15 other detached systems with M dwarf companions
that have orbital information and masses for both the white dwarf and the 
companion that could be analyzed in the same way \citep{ritter03}.  Medium to high resolution 
optical spectra could be used to find out how many of these objects possess
metal lines, and whether the mass loss rates inferred for those systems match
the binaries in my sample.  

\acknowledgements
This work has made extensive use of both the Simbad and Vizier
services.  This paper has been greatly improved by an anonymous referee.
JD would like
to thank Steinn Sigurdsson and Alycia Weinberger for extremely helpful 
comments and insights.

\bibliography{g29bib}
\bibliographystyle{apj}
\clearpage
\begin{deluxetable}{ccccccc}
\tablecolumns{7}
\tablewidth{0pc}
\tablecaption{\label{tab:closeorb} Table of Close Binary Parameters}
\tablehead{
\colhead{WD} & \colhead{Period} & \colhead{Orbital Separation} & \colhead{M$_{WD}$} & \colhead{M$_{RD}$} & \colhead{R$_{RD}$} & \colhead{Reference} \\
& \colhead{(d)} & \colhead{(AU)} & \colhead{($\Msun$)} & \colhead{($\Msun$)} & \colhead{(R$_{\odot})$} & }
\startdata
0419-487 & 0.3037 & 7.3$\times$10$^{-3}$ & 0.47 & 0.095 & 0.189 & 1\\
1026+002 & 0.597259 & 0.013 & 0.68 & 0.23 & 0.25 & 2 \\
1213+528 & 0.667579 & 0.015 & 0.63 & 0.36 & 0.32 & 3 \\

\enddata
\tablerefs{(1) \citet{bruch99}, (2)\citet{saffer93}, (3)\citet{shimanskii02}}

\end{deluxetable}

\begin{deluxetable}{cccc}
\tablecolumns{4}
\tablewidth{0pc}
\tablecaption{\label{tab:wideorb} Table of Wide Binary Parameters}
\tablehead{\colhead{WD} & \colhead{Ang. Separation} & \colhead{Distance} &
\colhead{Orbital Separation} \\
& & \colhead{(pc)} & \colhead{(AU)}
}
\startdata
0354+463 & $<$ 0.2\arcsec & 30\tablenotemark{a} & 1.5$\pm$0.5 \\
1049+103 & 0.26\arcsec & 107\tablenotemark{b} & 28 \\
1210+464 & 1.04\arcsec & 153\tablenotemark{b} & 159 \\
\enddata
\tablenotetext{a}{\citet{sion}}
\tablenotetext{b}{\citet{liebert04}}
\end{deluxetable}

\begin{deluxetable}{ccccccc}
\tablecolumns{7}
\tablewidth{0pc}
\tablecaption{\label{resulttab} WD Parameters and Derived Mass Loss Rates}
\tablehead{\colhead{WD} & \colhead{T$_{eff}$} & \colhead{[Ca/H]} & 
\colhead{$\log{q}$} & \colhead{$\log{\Theta}$} & \colhead{$\dot{M}$} &
\colhead{$\dot{M}_{RD}$} \\
& \colhead{(K)} & & & \colhead{(yr)} & \colhead{($\Msun$/yr)} & \colhead{($\Msun$/yr)}
}
\startdata
0354+463 & 7765 & -8.335 & -8.8 & 3.37 & 6$\times$10$^{-16}$ & 6$\times$10$^{-10}$\\
0419-487 & 6296 & -9.284 & -7 & 4.5 & 4$\times$10$^{-16}$ & 6$\times$10$^{-15}$ \\
1026+002 & 14300 & -8.56 & -16.6 & -2 & 2$\times 10^{-18}$ & 1$\times 10^{-16}$ \\
1049+103 & 19800 & -7.182 & $<$-18 & $<$-2.2 & 3$\times$10$^{-18}$ & 7$\times$10$^{-10}$ \\
1210+464 & 27000 & -7 & $<$-18 & $<$-2.2 & 4$\times$10$^{-18}$ & 5$\times$10$^{-8}$ \\
1213+528 & 13000 & -8.1 & -16 & -0.6 & 9$\times$10$^{-19}$ & 1$\times$10$^{-16}$ \\
\hline
\hline
Proxima Centauri & & & & & & 4$\times 10^{-15}$,6$\times 10^{-14}$ \\
\enddata
\end{deluxetable}
\clearpage

\end{document}